\documentclass[12pt]{article}
\usepackage{amssymb}
\usepackage{amsmath}
\usepackage{amsfonts}
\usepackage{epsfig}
\usepackage{anysize}
\marginsize{2cm}{2cm}{2.5cm}{3cm}
\date{}

\begin{document}

\title{\LARGE \bf  Scaling of distributions of sums of positions for chaotic dynamics at band-splitting points}

\author{Alvaro D\'{i}az-Ruelas\textsuperscript{1}, Miguel Angel Fuentes\textsuperscript{2,3,4}, Alberto Robledo\textsuperscript{1}\\ 
\footnotesize 1. Instituto de F\'{i}sica, Universidad Nacional Aut\' onoma de M\' exico y Centro de Ciencias de la Complejidad, \\ 
\footnotesize Apartado Postal 20-364, M\'exico 01000 DF, Mexico.\\
\footnotesize 2. Santa Fe Institute, 1399 Hyde Park Road, Santa Fe, New Mexico 87501, USA\\
\footnotesize 3. Instituto de Investigaciones Filos\' oficas, SADAF, CONICET, Bulnes 642, 1428 Buenos Aires, Argentina\\
\footnotesize 4. Instituto de Sistemas Complejos de Valpara\'iso, Subida Artillería 470, Valpara\'iso, Chile\\
       }
			
\maketitle




\abstract{The stationary distributions of sums of positions of trajectories generated by the logistic map have been found to follow a basic renormalization group (RG) structure: a nontrivial fixed-point multi-scale distribution at the period-doubling onset of chaos and a Gaussian trivial fixed-point distribution for all chaotic attractors. Here we describe in detail the crossover distributions that can be generated at chaotic band-splitting points that mediate between the aforementioned fixed-point distributions. Self affinity in the chaotic region imprints scaling features to the crossover distributions along the sequence of band splitting points. The trajectories that give rise to these distributions are governed first by the sequential formation of phase-space gaps when, initially uniformly-distributed, sets of trajectories evolve towards the chaotic band attractors. Subsequently, the summation of positions of trajectories already within the chaotic bands closes those gaps. The possible shapes of the resultant distributions depend crucially on the disposal of sets of early positions in the sums and the stoppage of the number of terms retained in them.}

\vspace*{10pt}

\noindent {\tiny PACS \ $5.45.$Ac Low-dimensional chaos\\
PACS \ $05.45.$Pq Numerical simulations of chaotic systems\\
PACS $05.10.$Cc \ Renormalization group methods}\\

\section{Introduction}

A few years ago \cite{tsallis1} a possible generalization of the central limit theorem (CLT) was put forward, as suitable for strongly correlated variables and that would have as its stationary distribution the so-called $q$-gaussian function \cite{tsallis1}. Subsequently, it was surmised that a fitting model system for the observation of this generalization would be the period-doubling accumulation point of the logistic map \cite{tbt1}. This development led to increased interest and discussion \cite{tbt2}-\cite{fuentes3} about whether sums of correlated deterministic variables at vanishing, or near vanishing, Lyapunov exponent $\lambda $ give rise to a general type of non-gaussian
stationary distribution.

As it turned out \cite{tbt2}, \cite{afsar1}, \cite{afsar2}, the distributions resembling $q$-gaussians at the period-doubling accumulation point require, unusual, specific procedures to be obtained. The first one is to work with a small but positive Lyapunov exponent $\lambda \gtrsim 0$. The
second is to discard an initial tract of consecutive positions in the ensemble dynamics, the disposal of a `transient', before evaluating the sum of the remaining positions. And the third is to stop the summation at a finite number of terms. When the transient set of terms is not discarded the resulting distribution would show an irregular, jagged, serrated, shape, whereas if the summation continues towards a larger and larger total number of terms the distribution approaches a gaussian shape. The $q$-gaussian-like distributions were observed along a sequence of values of the map control parameter $\mu $ that in latter studies \cite{afsar2} were identified as those approximately obeying the Huberman-Rudnick scaling law \cite{huberman1}, the power law that relates distance in control parameter space to Feigenbaum's universal constant $\delta $, or, equivalently, the number $2^{n}$, $n=0,1,2,\ldots $, of bands of the chaotic attractors.

\begin{figure*}[t] 
\centering
\epsfig{file=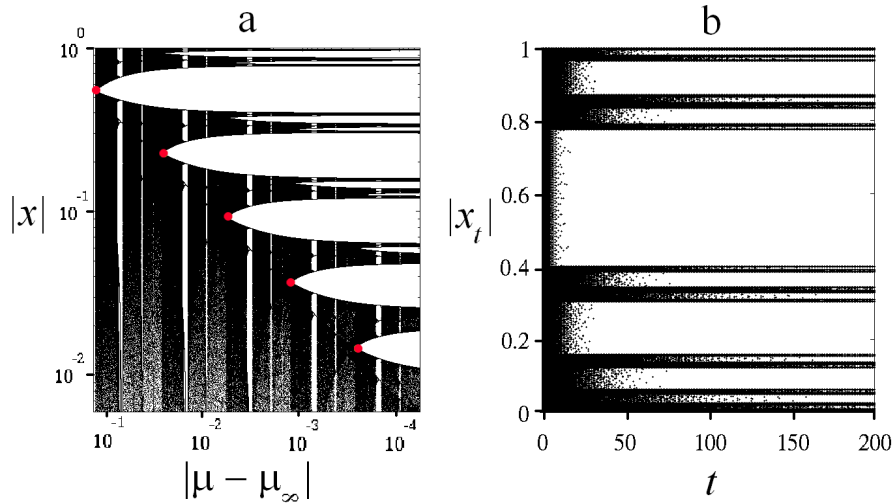, width=.74\textwidth} 
\caption{ \scriptsize (a) Attractor bands (in black) and gaps between them (white horizontal regions) in logarithmic scales, $-\log (|\mu-\mu_{\infty}|)$ and $\log (|x|)$ in the horizontal and vertical axes, respectively. The band-splitting points $M_n$ (circles) follow a straight line indicative of power law scaling. The vertical white strips are periodic attractor windows. (b) Sequential gap formation for $M_{5}$ by an ensemble of trajectories with initial conditions uniformly distributed along the map phase space. Black dots represent absolute values of trajectory positions $|x_t|$ at iteration time $t$. See text.}
\label{Fig:Fig1}
\end{figure*}

Here we provide a thorough rationalization, backed by ample evidence, of the properties of sums of consecutive positions and their distributions for ensembles of trajectories associated with the sequence of chaotic $2^{n}$-band attractors of the logistic map. We add to previous understanding \cite{fuentes1}-\cite{fuentes3} on the distributions of sums of positions at the period-doubling accumulation point for trajectories initiated within the attractor or with an ensemble of them uniformly-distributed across the entire phase space (the domain of the map). In the former case \cite{fuentes1},  \cite{fuentes2} the support of the stationary distribution is the multifractal set that makes up the Feigenbaum attractor and its amplitude follows its multifractal nature. For the latter case \cite{fuentes3} we demonstrated that the stationary distribution possesses an infinite-level hierarchical structure that originates from the properties of the repellor set and its preimages. We have also established \cite{fuentes1}, \cite{fuentes2} that the entire problem $\lambda $ $\geq 0$ can be couched
in the language of the renormalization group RG formalism in a way that makes clear the identification of the existing stationary distributions and the manner in which they are reached. The RG transformation consists of position summation (and rescaling); there is only one relevant variable, the control parameter distance to the transition to chaos $\Delta \mu $. There are two fixed-point distributions, the trivial continuum-space gaussian distribution and the nontrivial multiscale distribution reached only when $\Delta \mu =0$%
. The RG transformation modifies behavior similar to that of the nontrivial fixed point into that resembling the trivial fixed point through a well-defined crossover phenomenon. We show here that it is at this crossover region that the $q$-gaussian-like distributions are observed in Refs. \cite{tbt1}-\cite{afsar2}.

In the following Section 2 we set up the elements of our analysis: The chaotic band splitting cascade of the logistic map \cite{schuster1}, \cite{beck1}, along which we study trajectories at the control parameter points where bands split, also called Misiurewicz ($M_{n}$) points \cite{beck1}. We focus on scaling properties for the sequence of $M_{n}$ points. There we explain the dynamics undergone by an ensemble of uniformly distributed initial positions that consists of consecutive gap formation until arrival at the $M_{n}$ attractor, after which intraband chaotic motion drives the dynamics. In Section 3 we present summations of positions and their distributions at various $M_{n}$ points for different choices of disposal of initial sets of positions and different total number of summation terms. We explain the structure of the sums and their distributions in terms of the dynamics described in Section 2. In particular we detail the case that leads to distributions that resemble a $q$-gaussian shape. In Section 4 we discuss our results at some length in terms of the associated RG transformation.

\section{Dynamics at chaotic band splitting points}

We consider the logistic map $f_{\mu }(x)=1-\mu x^{2}$, $-1\leq x\leq 1$, $0\leq \mu \leq 2$, for which the control parameter value for its main period-doubling cascade accumulation point is $\mu =$ $\mu _{\infty}=1.401155189092..$. When $\mu $ is shifted to values larger than $\mu _{\infty }$, $\Delta \mu \equiv \mu -\mu _{\infty }>0$, the attractors are (mostly) chaotic and consist of $2^{n}$ bands, $n=0,1,2,...$, where $2^{n}\sim \Delta \mu ^{-\kappa }$, $\kappa =\ln 2/\ln \delta $, and $\delta =4.669201609102 \ldots$ is the universal constant that measures both the rate of convergence of the values of $\mu =\mu _{n}$ to $\mu _{\infty }$ at period doubling or at band splitting points. See Fig. 1a. The Misiurewicz ($M_{n}$) points, are attractor merging crises, where multiple pieces of an attractor merge together at the position of an unstable periodic orbit  \cite{grebogi1}. The  $M_{n}$ points can be determined by evaluation of the trajectories with initial condition $x_{0}=0$ for different values of $\mu $,\ as these orbits follow the edges of the chaotic bands until at $\mu =\mu _{n}$ the unstable orbit of period $2^{n}$ reaches the merging crises \cite{grebogi1}. \

Trajectories initiated inside a $2^{n}$-band attractor consist of an interband periodic motion of period $2^{n}$ and an intraband chaotic motion. Trajectories initiated outside a $2^{n}$-band attractor exit progressively a family of sets of gaps formed in phase space between the $2^{n}$ bands. This family of sets of gaps starts with the largest gap formed around the first unstable orbit, or first repellor, of period $2^{0}$, followed by two gaps containing the two positions of the second repellor of period $2^{1}$, and so on. See Fig. 1b. The widths of the gaps diminish in a power law fashion as their numbers $2^{k}$, $k=0,1,2,...$, for each set increase. We follow the dynamics towards the $M_{n}$, $n=0,1,2,...$, attractors by setting a uniformly-distributed ensemble of initial conditions across phase space, $-1\leq x_{0}\leq 1$, and record the normalized number of bins $W_{t}$, in a fine partition of this interval, that still contain trajectories at iteration time $t$. The results are shown in Fig. 2, where we observe an initial power law decay in $W_{t}$ with logarithmic oscillations followed by a transition into a stay regime, a plateau with a fixed value of $W_{t}$, when (practically) all trajectories become contained and remain in the bands of the attractor.

The properties of $W_{t}$ show discrete scale invariance associated with powers of $2$ characteristic of unimodal maps. The number of logarithmic oscillations in the regime when trajectories flow towards the attractor coincides with the number of consecutive sets of gaps that need to be formed at the \ $M_{n}$\ points, whereas the final constant level of $W_{t}$ coincides with the total number of bins that comprise the total width of the $2^{n}$ bands of the attractors. We notice that these properties when observed along the plateau entry points labeled $t^{*}_{n}$ shown in Fig. 2 obey the Huberman-Rudnick scaling law since the times $t^{*}_{n} $ are related to the $2^{n}$ bands of the $M_{n}$ points and these in turn are
given by $\Delta \mu _{n}\sim \delta ^{-n}$.

\begin{figure} 
\centering
\epsfig{file=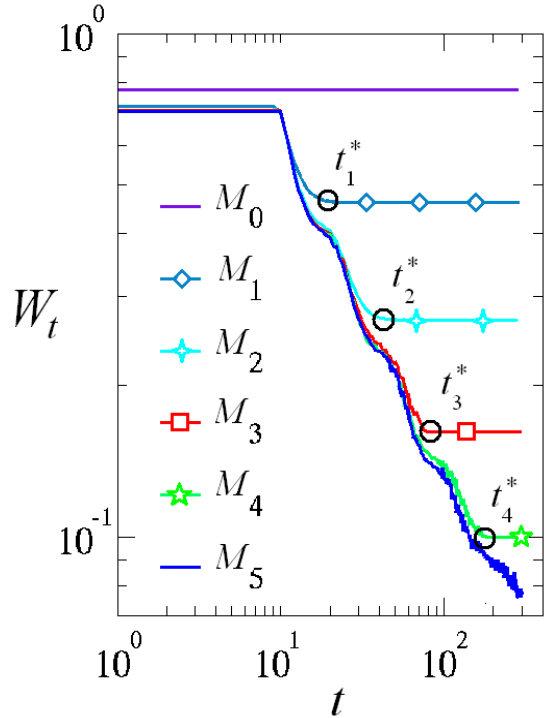, width=.48\textwidth}
\caption{\scriptsize Normalized number $W_t$ of bins containing trajectories at iteration time $t$ in logarithmic scales. A uniform distribution of $10^6$ initial conditions across $[-1,1]$ were placed in a partition of $10^6$ bins. The curves shown correspond to the band-splitting points $M_n, \ n=0,1,\ldots,5$. For each $n>0$ there is an initial power-law decay with logarithmic oscillations followed by a final constant plateau. The former corresponds to sequential gap formation and the latter indicates that all trajectories are within the attractor bands. The circles labeled $t^{*}_{n}$ indicate the plateau entry times. } \label{Fig:Wt}
\end{figure}

\section{Sums of positions and their distributions at band splitting points}

\begin{figure}
\centering
\epsfig{file=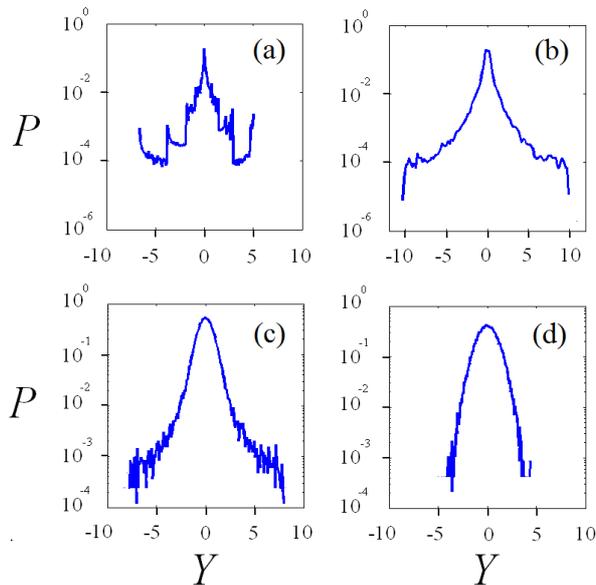, width=.48\textwidth}
\caption{\scriptsize Distributions $P(Y;N_{s},N_{f};\mu _{n})$ of centered sums $Y \equiv X - \langle X \rangle$, where $X$ is given by Eq. (\ref{Eq:sums}). The sums were obtained from a uniform distribution of $10^6$ initial conditions across $[-1,1]$ at $M_5$ when the attractor of $2^5$ bands is about to split into $2^6$ bands. A value of $N_s = 2^8$ is used in all panels. The values of $N_f$ used are: (a) $2^{5}$, (b) $2^{9}$, (c) $2^{13}$ and (d) $2^{17}$. See text.} \label{Fig:DistributionsM_X}
\end{figure}

We consider now the sum of consecutive positions $x_{t}$ starting with an iteration time $t = N_{s}$ up to a final iteration time $t = N_{s}+N_{f}$ of a trajectory with initial condition $x_{0}$ and control parameter value $\mu $ fixed at an $M_{n}$\ point, $n=0,1,2,...$, \textit{i.e.}

\begin{equation}
X(x_{0},N_{s},N_{f};\mu _{n})\equiv \sum\limits_{t=N_{s}}^{N_{s}+N_{f}}x_{t}.
\label{Eq:sums}
\end{equation}

We studied a collection of these sums for trajectories started from a uniform distribution of initial conditions in the entire interval $-1\leq x_{0}\leq 1$ with different values of $n$, $N_{s}$ and $N_{f}$, and we also evaluated their corresponding histograms and finally their distributions by centering and normalization of the histograms. Clearly, stationary distributions require $N_{f}\rightarrow \infty $ and, unless there is some unusual circumstance, they are not dependent on the value of $N_{s}$. We
know \cite{fuentes1}, \cite{fuentes2} that for all chaotic attractors ($\Delta \mu >0$) the stationary distribution is gaussian and that in the limit $\Delta \mu =0$ the stationary distribution is of an exceptional kind with intricate multiscale features \cite{grassberger1}, \cite{fuentes3}. Here we explore other distributions that can be obtained when $N_{s}$ and $N_{f}$ are varied and identify the dynamical properties that give rise to them.

The observation of $q$-gaussian-like distributions in Refs. \cite{tbt2}, 
\cite{afsar2} involved a large value of discarded terms $N_{s}$ before sums
similar to that in Eq. (\ref{Eq:sums}) were evaluated. Also, it was found
necessary to limit the number of summands to a finite number $N_{f}$ to
prevent the distribution approach a gaussian form. For example in Ref. 
\cite{tbt2} a fixed value of $N_{s}=2^{12}$ was reported to be used for sums
evaluated at attractors with a number of bands $2^{n}$ with $n$ in the range 
$4$ to $8$. These sums were terminated, respectively, with values $N_{f}=2^{n_{f}}$
with $n_{f}$ in the range $9$ to $17.$ In these studies the values of $
\Delta \mu $ were not precisely fixed at band splitting points as we do here
but the dynamical properties we describe are equivalent. We can
understand the effect of the values of $n$, $N_{s}$ and $N_{f}$ used in terms of the
dynamics of trajectories from the knowledge gained in the previous section. In references \cite{tbt2} and \cite{afsar2}, the starting times $t = N_{s}$ in the sums in Eq. \ref{Eq:sums} satisfy the condition $t^{*}_{n} \ll N_{s}$. We can conclude with the assistance of Fig.2, that the terms discarded in those studies comprise the flow of trajectories towards the attractors plus a significant segment of dynamics within the chaotic bands, therefore all of the terms contained in the sums correspond to the dynamics within the chaotic bands.

As a representative example we show in Fig. 3 the distributions $P(Y;N_{s},N_{f};\mu _{n})$  for the sums in Eq. (\ref{Eq:sums}), with $Y = X - \langle X \rangle$, and where $\langle X \rangle$ is the average of $X$ over $x_{0}$. In this figure $n=5$ and $N_{s}=2^{8}$, and $N_{f}$ takes the values $N_{f}=2^{5}$, $2^{9}$, $ 2^{13}$ and , $2^{17}$, respectively, in panels (a), (b), (c) and (d). In (a) the sum comprises only one visit to each band and the structure of the distribution is the result of one cycle intraband motion of the ensemble of trajectories. In (b) the sum contains already about $2^{4}=16$, band cycles, for which we obtain a distribution with $q$-gaussian-like
shape but sharp drops at the edges. In (c) the $q$-gaussian-like shape is disappearing after $256$ band cycles, while in (d), when there are $4096$ band cycles, we observe already the stationary gaussian form. The same distribution progression pattern shown in Fig. 3 is observed at other $M_{n}$ points. Furthermore, the sums and their distributions for any value of $n$ can be reproduced by rescaling consistent with Huberman-Rudnick law. This is illustrated in Fig. 4 where we show in panels (a), (b) and (c) the resemblance of the centered sums $Y$ for the band merging points $M_3, \ M_4$ and $M_5$, respectively. In panel (d) we show the distributions $P$ for these sums without rescaling of the horizontal axis $Y$.

\section{Summary and discussion}

\begin{figure*}
\centering
\epsfig{file=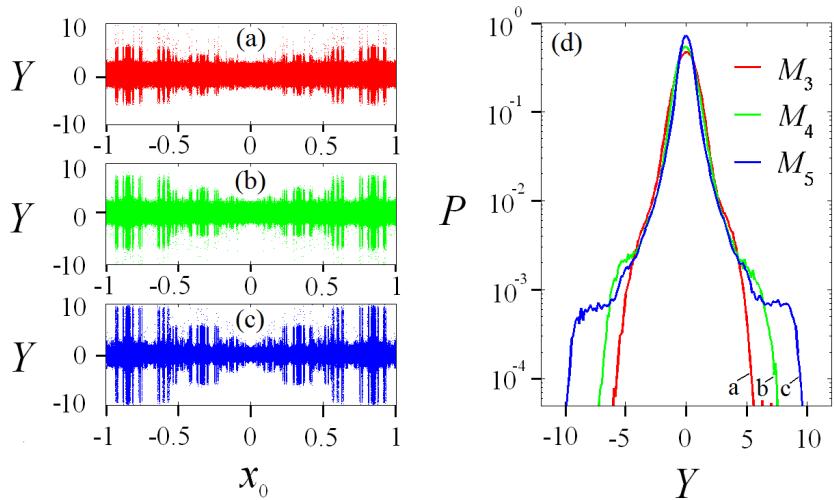, width=.7\textwidth}
\caption{\scriptsize Rescaled sums obtained from a uniform distribution of $3*10^6$ initial conditions across $[-1,1]$ at the band-splitting points $M_n, \ n= 3,4,5 $ with labels (a),(b),(c), respectively, and with their corresponding distributions in (d). The values of $N_s$ and $N_f$  used are, respectively $N_s = 2^6,2^7,2^8,$ and  $N_f = 2^{12}, 2^{13}, 2^{14}$.}
 \label{Fig:Sums_and_Dist}
\end{figure*}

We have shown that there is an ample variety of distributions $P(Y;N_{s},N_{f};\mu _{n})$ associated with the family of sums of iterated positions, as in Eq. (\ref{Eq:sums}), obtained from an ensemble of trajectories started from a uniform distribution of initial conditions in the interval $-1\leq x_{0}\leq 1$. The shapes of these distribution vary with $N_{s}$ and $N_{f}$ but there is scaling property with respect to $n$. All the types of distributions obtained can be understood from the knowledge of
the dynamics that these trajectories follow, both when flowing towards the chaotic band attractors and when already within these attractors. There exists throughout the family of chaotic band attractors with $\lambda>0$ an underlying scaling property, displayed, e.g., by the self-affine
structure in Fig. 1a. This scaling property is present all over, here
highlighted by: i) The sequential formation of gaps shown in Fig. 1b. ii)
The number of bins $W_{t}$ still containing trajectories
at iteration time $t$, shown in Fig. \ref{Fig:Wt}, both for its initial decay with logarithmic
oscillations and the final constant regime. And iii) the different classes of sums and their distributions obtained for a given value of $n$ are reproduced for other values of $n$ under appropriate rescaling, as shown in Fig. \ref{Fig:Sums_and_Dist}. For adeptness and precision purposes we chose here to study the family of Misiurewicz points $M_{n}$ but similar, equivalent, results are obtained
for chaotic attractors between these points.

The discussion about the types of distributions $P(Y;N_{s},N_{f};\mu
_{n})$ is assisted by recalling \cite{fuentes1}, \cite{fuentes2} the RG
framework associated with summation of positions. Positions $x_{t}$ for
trajectories within chaotic-band attractors can be decomposed as $x_{t}=$ $%
\overline{x}_{t}+\delta x_{t}$, where $\overline{x}_{t}$ is chosen to be
(for example) fixed at the center of the band visited at time $t$ and $\delta x_{t}$
is the distance of $x_{t}$ from $\overline{x}_{t}$. When the number of bands 
$2^{n}$ is large all the values of $\delta x_{t}$ are small. The sum in Eq. (
\ref{Eq:sums}) can be written as

\begin{equation}
X\equiv \overline{X}+\delta X,\;\overline{X}=\sum\limits_{t=N_{s}}^{N_{f}}
\overline{x}_{t},\;\delta X=\sum\limits_{t=N_{s}}^{N_{f}}\delta x_{t},
\label{Eq:sum2}
\end{equation}

\noindent where $\overline{X}$ captures the interband periodic (and therefore correlated) motion and $\delta X$ consists of the intraband chaotic (and therefore random) motion. As discussed in Refs. \cite{fuentes1}, \cite%
{fuentes2} the action of the RG transformation, summation, is driven by $%
\delta x_{t}$ and results in gradual widening of all the chaotic bands, such
that eventually for a sufficiently large number of summands all of them
merge into a single band. When $0\leq N_{s}\lesssim t^{\ast }_{n}$ gap formation
competes with band widening, while when $t^{\ast }_{} \lesssim N_{s}$ band
widening develops unimpeded. When $0\leq N_{s} \lesssim t^{\ast }_{n}$ the
combined processes of the dynamical evolution of the ensemble of
trajectories and the repeated RG transformation is dominated initially by gap formation but it is
always followed by gap merging. Initially, the distributions for these sums
resemble the jagged multiscale shape of the stationary distribution for the
nontrivial fixed point at $\Delta \mu =0$ but they necessarily evolve
towards the gaussian distribution of the trivial fixed point present for $\Delta \mu >0$ \cite{fuentes1}, \cite{fuentes2}. When $t^{\ast }_{n} \ll N_{s}$,
as in Refs. \cite{tbt1}, \cite{tbt2}, \cite{afsar1}, \cite{afsar2}, the
trajectory positions considered in the sums are all contained within the
attractor bands and from the first term $t = N_{s}$ the gaps begin to close
due to the action of $\delta x_{t}$ that is akin to an independent random
variable. As we have shown in Fig. 3, when the number of summands grow the
shape of the distribution evolves by first eliminating the initial serrated
features, then developing a symmetrical shape that shows possible long
tails but that end in a sharp drop (the claimed $q$-gaussian type), and finally
the approach to the gaussian stationary distribution. All of the above can
be observed for each $2^{n}$-band chaotic attractor, basically from $n\geq 1$
, and when a self-affine family of these attractors is chosen, like the
Misiurewicz points $M_{n}$ the sums and their distributions can be rescaled
such that they just about match for all $n$, as shown in Fig. \ref{Fig:Sums_and_Dist}, where the sums where started at $N_{s} \simeq t^{*}_{n}$.

Concisely, the elimination of a large enough set of early positions in the sums for a given $n$, such that the location of its first term $N_{s}$ is located inside the plateau of $W_{t}$ in Fig. 2, ensures that the sums capture only the dynamics within the $2^{n}$-band attractor. Therefore the
shape of the distributions are dominated by the uncorrelated chaotic contributions $\delta x_{t}$, that as $t$ increases evolves towards the final gaussian shape. A nongaussian distribution can only be obtained if there is\ a finite number of summands $N_{f}$. Self-affinity in the chaotic-band family of attractors, provides scaling properties to the distributions of sums of positions that are described by an appropriate use of the Huberman-Rudnick power-law expression. 


\section*{Aknowledgements} MAF thanks CONICTY Project: Anillo en Complejidad Social SOC-1101, FONDECYT 1140278. AR and AD-R acknowledges support from DGAPA-UNAM-IN103814 and CONACyT-CB-2011-167978 (Mexican Agencies).

\end{document}